\documentclass[letterpaper,twocolumn,10pt]{article}
\usepackage{graphicx} % Required for inserting images
\usepackage{multicol}
\usepackage{anyfontsize}
\usepackage{array}
\usepackage{url}
\usepackage{hyperref}
\usepackage[version=4]{mhchem}

\begin{document}

%don't want date printed
\date{}

%make title bold and 14 pt font (Latex default is non-bold, 16 pt)
\title{\Large \bf ICT Sector Greenhouse Gas Emissions - Issues and Trends}

\author{
{\rm Peter Garraghan}\\
School of Computing\\ \& Communications\\
Lancaster University
\and
{\rm John Hutchinson}\\
School of Computing\\ \& Communications\\
Lancaster University
\and
{\rm Adrian Friday}\\
School of Computing\\ \& Communications\\
Lancaster University
}

\maketitle

% Use the following at camera-ready time to suppress page numbers.
% Comment it out when you first submit the paper for review.
\thispagestyle{empty}

\begin{abstract}
As Information and Communication Technology (ICT) use has become more prevalent, there has been a growing concern in how its associated greenhouse gas emissions will impact the climate. Estimating such ICT emissions is a difficult undertaking due to its complexity, its rapidly changing nature, and the lack of accurate and up-to-date data on individual stakeholder emissions. In this paper we provide a framework for estimating ICT's carbon footprint and identify some of the issues that impede the task. We attempt to gain greater insight into the factors affecting the ICT sector by drawing on a number of interviews with industry experts. We conclude that more accurate emissions estimates will only be possible with a more more detailed, industry informed, understanding of the whole ICT landscape and much more transparent reporting of energy usage and emissions data by ICT stakeholders.
\end{abstract}

\section{Introduction}
The role of Information and Communication Technology (ICT) is increasingly important and affects all activities from entertainment, business, government and beyond. Huth characterises ICT systems as ``an umbrella term pertaining to any digital computer and communication devices spanning radio, television, cellular phones, computer hardware and software, and satellite systems'' \cite{HUTH2017131}. ICT is now so prevalent in daily life that there is a growing awareness that its provision comes with a significant carbon footprint (estimated to be between 1.8 and 3.9\% of global emissions \cite{FREITAG2021100340}). The physical components that enable ICT include networking equipment, mobile networks, legacy landlines, data centres, network attached sensors and actuators, and a range of end-user devices used to both create and consume data (e.g., laptops, smartphones, tablets and TVs). In the UK, such ICT systems are leveraged to deliver a wide variety of digital services underpinning the economy and thus contribute to the national carbon footprint.
  
Measuring the size of the ICT carbon footprint is challenging as it requires estimations to be made of the greenhouse gas (GHG) emissions associated with all phases of the ICT life cycle from manufacturing to usage and disposal. All of these values are constantly changing and, even when monitored and measured, many are not openly declared. Understanding the impact of these emissions is further complicated given that ICT adoption across other industry sectors is thought to reduce their respective carbon footprint from extensive digitisation of business processes.  This is an argument that is often used, but seldom evidenced to shift focus away from ICT industry and its responsibilities in terms of environmental impact.

The fast-changing nature of ICT, the growth of ICT infrastructure, its relation to a changing energy system and the complex supply chains all impact organisations’ ability to make informed decisions that could lead to a reduction in GHG emissions.

The long term goal of this work is to create a detailed model of the whole ICT landscape which could be used to help us identify all of the sector's stakeholders and their roles and responsibilities, and also to reason about the associated GHG emissions. In this paper, we model of the ICT landscape specifically associated with video streaming and we then use that as the basis for exploring what might be driving emissions growth by eliciting the views of industry experts.  Globally, in 2022 video streaming accounted for more than 65\% of Internet traffic \cite{sandvine2023}, making it a particularly significant part of ICT use.

In section 2, we outline the methods used to conduct our study. In section 3, we provide background work on measuring the ICT GHG emissions, present our landscape model and describe a process for estimating the emissions associated with video streaming. We explore the views of industry experts in section 4. In section 5, we discuss our findings and section 6 provides a brief conclusions and thoughts about future work.
I wish you the best of success.

\section{Methodology}

We adopted different methodologies for the different parts of our study. The model of the ICT landscape was used a as a basis for the qualitative research with industry stakeholders.

\subsection{ICT Landscape}

As part of our longer term goal of creating a complete model of the ICT landscape, we chose to focus on the landscape as it now exists to support video streaming. To achieve this we combined literature based analysis with specialist knowledge of industry experts.

When conducting the review of the available literature, we observed that studies would model ICT systems at differing levels of abstraction (CPU, server, facility, national-level). To ensure that all components were successfully mapped to the same abstraction level, we conducted a series of interviews and workshops (incorporating external feedback from various ICT stakeholders within the networking and telecommunications industry) to refine the scope of the ICT landscape in an iterative manner. This decision was taken to ensure adequate separations of concern and to identify boundaries to reduce complexity (i.e., modelling every single sub-component within the ICT landscape at the processor-level resulted in an unwieldy and convoluted landscape diagram). 

The decision to specifically target video streaming was taken to define a clear analysis boundary, as well as take advantage of expected data availability/industry maturity within the UK. The ICT landscape created focuses primarily within the Use phase (e.g., the operation, transmission, and consumption of digital content).

\subsection{Industry Stakeholders}

\begin{figure}[!h]
    \centering
    \includegraphics[width=8cm]{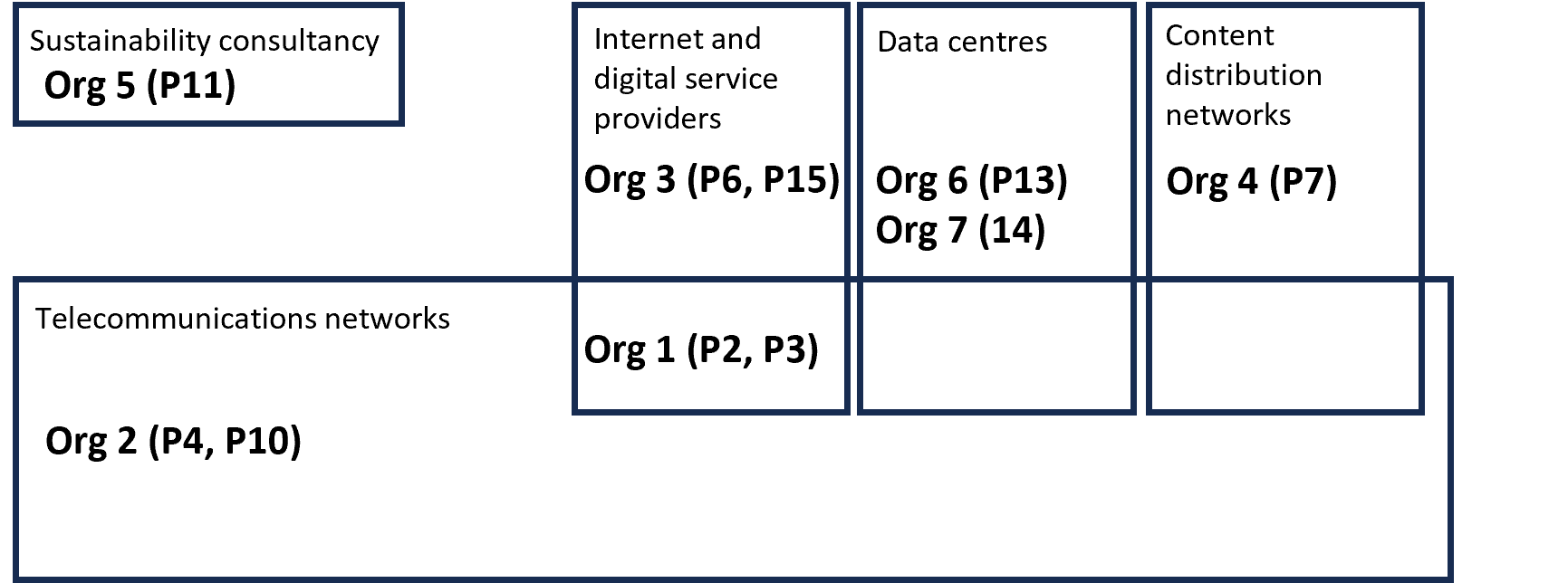}
    \caption{Map of stakeholder groups, organisations and participants interviewed.}
    \label{fig:partic}
\end{figure}

We conducted 10 semi-structured interviews with representatives from digital service providers, telecommunications companies, data centre consultants, and research consultancies.  These representatives were chosen as being members of stakeholder groups in the video streaming supply chain within the UK identified as relevant \cite{10.1145/3290605.3300627}. Some of these organisations spanned more than one identified stakeholder group (i.e., telecommunications and Internet or other digital services). We also enlisted participants with leading roles in the organisations from a variety of functions (research and development and operations management) and different areas (i.e., sustainability, network, content delivery, distribution, innovation), to gain a diversity of perspectives.  Figure \ref{fig:partic} illustrates how the interviewees relate to the video streaming infrastructure.  Table \ref{table:participants} lists the participants by identifier and role.

\begin{table*}[]
\fontsize{9pt}{9pt}\selectfont
\centering
\caption{Interview participants by stakeholder group.}\begin{tabular}{lllll}
\hline
\textbf{ID} & \textbf{Role}   & \textbf{Area}           & \textbf{Stakeholder Group}  & \textbf{Company} \\ \hline
P2                      & Head            & Content                 & Digital Service Provider    & 1                     \\
P3                      & Head            & Video                   & Digital Service Provider    & 1                     \\
P4                      & Head            & Sustainability          & Telecoms                    & 2                     \\
P6                      & Head            & Distribution            & Digital Service Provider    & 3                     \\
P7                      & Founder         & CDN Expert              & Content Delivery Network    & 4                     \\
P10                     & Director        & Wireless Innovation     & Telecoms/Customer Equipment & 2                     \\
P11                     & Partner Manager & Sustainability          & Consultancy                 & 5                     \\
P13                     & Senior Partner  & Data Centre \& Cloud    & Data Centres                & 6                     \\
P14                     & Entrepreneur    & Data Centre Management  & Data Centres                & 7                     \\
P15                     & Lead Engineer   & Sustainable Engineering & Digital Service Provider    & 3                     \\ \hline
\end{tabular}
\label{table:participants}
\end{table*}

Our questions focused on participant’s role, organisation, relation to video streaming, as well as how decision-making relating to sustainability took place. The questions were tailored to each stakeholder group and participant role.  Interviews were carried out online and lasted for approximately 45-60 minutes. Interviews were recorded and transcribed by a third party before being thematically coded by the researchers. The interviews were not designed to collect data on the energy or GHG emissions, but focused on understanding the trends, drivers and barriers to the reduction of carbon emissions related to video streaming and its supporting ICT infrastructure. 

\section{ICT footprint}

The ICT carbon footprint measures the total greenhouse gas emissions caused directly and indirectly by a person, organisation, service or product. It is measured in tonnes or KG of carbon dioxide equivalent (\ce{CO2}e), combining the impact of different greenhouse gases into one figure equivalent to if it were all \ce{CO2}, based on their warming potential \cite{carbontrust2021}. For example, methane is a far more potent GHG with a global warming potential approximately 30 times that of \ce{CO2} over a 100-year timescale \cite{unece2023}.

The ICT carbon footprint is related to generating the energy required to energise ICT hardware to perform computation and power single devices to large-scale facilities such as data centres. All ICT devices require electricity to power different components (CPU/GPUs, memory, storage, network cards, displays, etc), which is predominantly converted into waste heat that, subsequently, requires removal via cooling (e.g., a computer fan, liquid cooling pumps). At a large scale, energy intensive cooling is required to ensure hardware operates within a given temperature range \cite{barroso2019datacenter}. But these emissions associated with the \textit{use} of ICT equipment and processes is only part of the overall picture.

A substantial, but less discussed aspect of ICT’s footprint is the emissions required during the manufacture, transport, recycling and disposal of the equipment, including within the extensive supply chains bringing together the raw materials. The embodied footprint of ICT is thus significant; for example, within a data centre, the energy intensity of computation means that embodied cost could be around 25 per cent of the lifetime emissions \cite{malmodin2020power}. For a mobile phone, this could be as much as 50 per cent, reinforcing the need to maximise a device’s use and usefulness for as long as possible \cite{malmodin2020power}. The total emissions of ICT should thus be understood in relation to its service operational lifetime, and the costs of its creation and disposal, as well as its use phase energy demand or operational efficiency.

\subsection{Issues affecting the accurate estimation of ICT GHG emissions}

Making accurate estimations of ICT GHG emissions is inherently difficult because of the scope of the task. However, it is further complicated by a number of issues that it should be possible to address if there was sufficient will to do so.

Many large companies publish annual reports to document their green credential (e.g., Google Environmental Report \cite{googer2022}) but are far from transparent about emissions and/or energy usage. For example, Google \cite{googer2022} typically reports the Power Usage Effectiveness (PUE) figures for their hyperscale data centres around the world. PUE is the ratio of a data centre's total energy usage to energy consumed by its IT operation. While such a figure is a useful measure of the efficiency of a data centre's operation, it says nothing about the amount of electricity consumed or the carbon intensity of that electricity. 

Further profound issues concerning the existing work on estimating the emissions associated with ICT are identified by Freitag et al. \cite{FREITAG2021100340}, namely the age of the date, the lack of ``interrogatability" of the data, potential conflicts of interest when some researchers are associated with ICT companies (and data is not available to others) and the lack of agreement about the scope of what constitutes ICT. 

\subsection{ICT Landscape for Video Streaming}

To date, measurement or estimation of the total UK ICT footprint covering wider aspects of the ICT landscape has yet to be completed \cite{POST2022}. We observed that such estimates can be found for global ICT emissions Malmodin and Lundén \cite{malmodin2018energy}, \cite{malmodin2018electricity}, Belkhir and Elmeligi \cite{BELKHIR2018448}, Andrae and Edler \cite{andrae2015global} at between 1.8\%–2.8\% (2.1\%–3.9\% when adjusting for supply chain pathways)\cite{FREITAG2021100340}. In this section, we will consider the process of estimating the emissions associated with video streaming as an example of ICT emissions in the UK.

To study the ICT footprint of video streaming, we created an ICT landscape that captures the core components of ICT operation within the UK. The ICT landscape was created via a review of academic publications and public emission reports capturing multiple ICT components or individual sub-components within a single ICT system.

\begin{figure*}[!ht]
    \centering
    \includegraphics[width=6in]{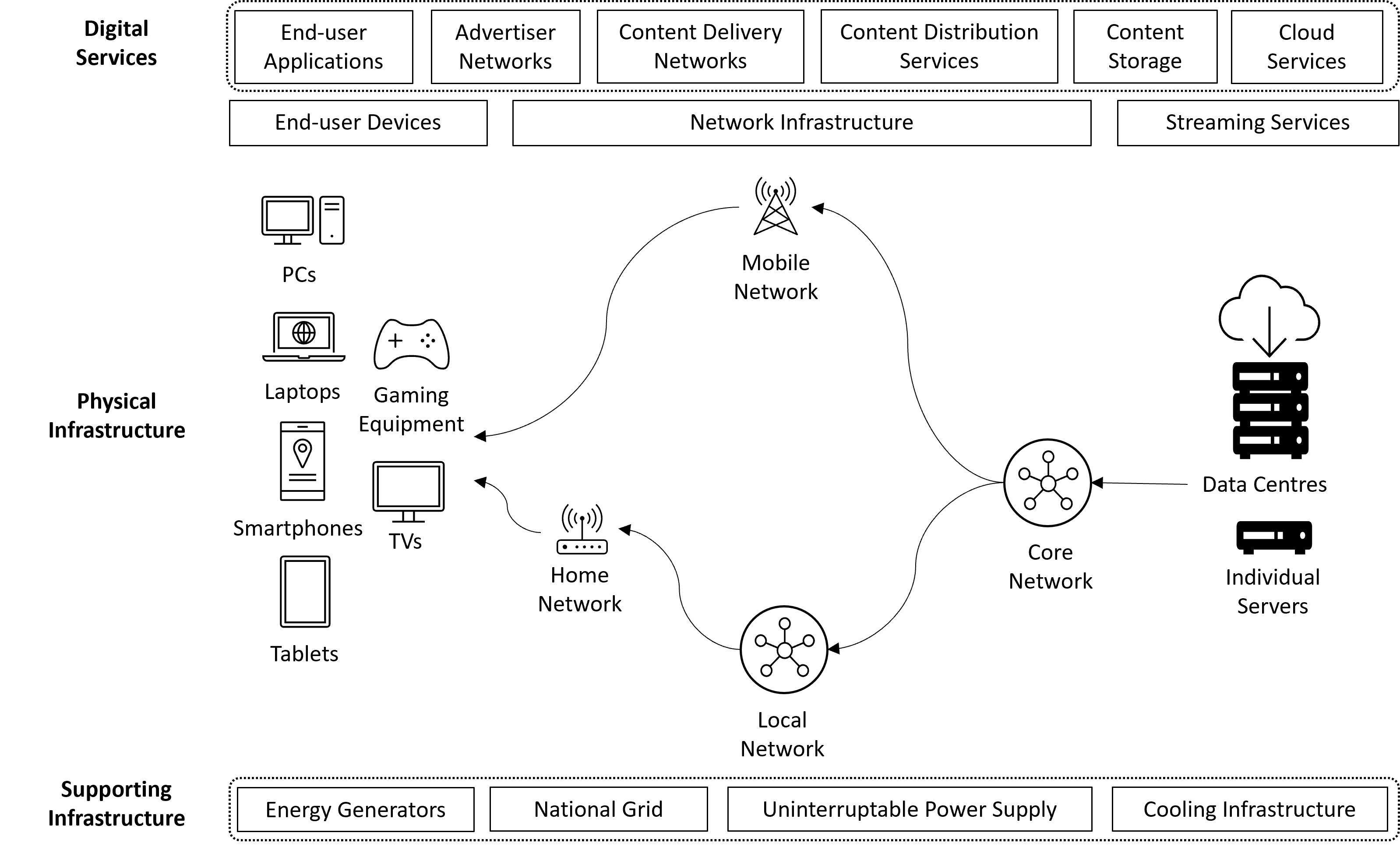}
    \caption{Simplified ICT landscape -–- use phase}
    \label{fig:landscape}
\end{figure*}

\subsection{ICT landscape mapping}

The ICT landscape as shown in Figure \ref{fig:landscape} can be viewed and measured from multiple perspectives. We found that the majority of existing ICT landscape mapping studies have been previously approached primarily from a mechanical perspective (i.e., measuring the footprint of physical systems) due to the ability to instrument power draw and carbon emissions. The physical infrastructure illustrated in Figure \ref{fig:landscape} maps closely onto that described by others (e.g., \cite{carbontrust2021}); we have additionally captured a broader set of perspectives including the energy and cooling infrastructure and the digital services that deliver content to users.

\subsubsection{Physical infrastructure}

Physical infrastructure entails the ICT system/device-level view of the ICT landscape for video streaming. At a high level, digital media content is transmitted from \textit{streaming services} through \textit{networking infrastructure}, and subsequently displayed on \textit{end-user devices} where it is consumed by users. It is worth noting that video streaming can be bi-directional (i.e., end-user devices are capable of transmitting media content through the network to streaming infrastructure that is then broadcast to other end-user devices (e.g., livestreaming, streaming of gameplay from PCs and consoles).

\begin{itemize}
  \item \textbf{Streaming Infrastructure}: ICT systems responsible for storing, transmitting, and servicing requests for video streaming. Streaming infrastructure comprises individual servers, server rooms, and data centres. At scale, streaming infrastructure can encompass hundreds, thousands, or even tens of thousands of co-located servers. Data centres are “compute powerplants”, and are facilities that host and consolidate compute, storage, and network systems to provision digital services efficiently \cite{cisco2020}. These facilities comprise a multitude of different services, software, hardware, networking, and cooling sub-components.
  \item \textbf{Network Infrastructure}: Video streaming is delivered to users via networked systems underpinning the Internet. This includes local area networks (LANs), wide area networks (WANs), transatlantic links, and the core Internet backbone. Collectively, these route video streaming data from the source to the end-user destination for consumption. These network components comprise routers, network switches, cables, long-range optical fibre cables and the equipment associate with the mobile data network.
  \item \textbf{End-user Devices}: To view requested digital content, a user can employ various devices such as laptops, smartphones, TVs, or games consoles. The electricity consumption of an individual end-user device is comparatively small in contrast to streaming and network infrastructure, yet on aggregate there are more than 286 million devices. The majority of the UK population own a smartphone (92\%), a laptop (78\%), or a smart TV (65\%) \cite{deloitte2022}.
\end{itemize}

These parts of the physical infrastructure is dependent upon \textbf{supporting infrastructure} ICT systems are predominantly maintained at an appropriate operating temperature using air or liquid cooling techniques. Whilst the total energy consumption of cooling within an end-user device (e.g., a computer fan) constitutes a marginal energy overhead compared with the CPU; at a large-scale provider the total footprint due to cooling can be up to 40\% of total data centre energy consumption \cite{nationalgrid2022}. The relationship between the energy for computation (useful work) and this overhead is the basis of the PUE metric described above. In addition to cooling, large scale facilities will often protect their service provision by the deployment of uninterruptible power supplies in the form of local power generation. This highlights the most fundamental supporting infrastructure, that of national energy generation and its distribution to users across the country.

The total of the physical infrastructure of the ICT landscape then facilitates multiple \textbf{digital services}, such as advertiser networks, cloud services and content delivery networks.

\subsection{Estimate of emissions associated with video streaming}

Using a methodology developed by DIMPACT \cite{dimpact2022}, the Carbon Trust produced an estimate of the European average GHG emissions associated with watching an hour of streamed video \cite{carbontrust2021}. The scope of their estimate is broadly the same as that shown in Figure \ref{fig:landscape} and omit the embedded carbon associated with that landscape. The details of the estimation process are described in depth in the white paper and the energy intensities of the different parts of the ICT landscape they used in their calculations were gleaned from various sources including direct measurements provided by DIMPACT members and estimates from the existing literature.

They estimate that the European average GHG emissions associated with an hour of video streaming is 56 g\ce{CO2}e. This is calculated using a European average grid intensity for the average energy consumption of 188 Wh per hour of video streaming. The breakdown of emissions and energy usage for the different components of the ICT landscape are given in Table \ref{table:breakdown}.

\begin{table*}[t]
\fontsize{9pt}{9pt}\selectfont
\centering
\caption{Breakdown of emissions and energy usage.}
\begin{tabular}{llll}
\hline
\textbf{Component}            & \textbf{Emissions}  & \textbf{Energy}  & \textbf{\% of total} \\ \hline
Data centres (including CDNs) & 1 g\ce{CO2}e/hour  & 1 Wh/hour  & 1\%         \\
Network transmission          & 6 g\ce{CO2}e/hour  & 20 Wh/hour & 10\%        \\
Home routers                  & 21 g\ce{CO2}e/hour & 71 Wh/hour & 38\%        \\
End-user device               & 28 g\ce{CO2}e/hour & 96 Wh/hour & 51\%        \\ \hline
\end{tabular}%
\label{table:breakdown}
\end{table*}

56 g\ce{CO2}e/hour of video streaming is the estimate of the European average but the actual emissions associated with video streaming in each country vary considerably according to the carbon intensity of the local electrical grid. For example, in the Carbon Trust's analysis, the estimate of GHG emissions for Sweden is 3 g\ce{CO2}e/hour of video streaming and that for Germany is 76 g\ce{CO2}e/hour.

The large variations in carbon intensity of the local electrical grid for different countries (in 2022 ranging from 29 g\ce{CO2}e/kWh for Norway to 767 g\ce{CO2}e/kWh for Kosovo \cite{wid2023}) will likely play a greater role in the emissions associated with ICT than the relative efficiencies of the ICT infrastructure. However, since average worldwide Internet penetration is around 66\% (as of January 2024 \cite{stat2024}), many countries with a high carbon intensity of their electrical grid will expect to see increased ICT usage going forward and so ICT infrastructure efficiency remains important.

\section{Stakeholder perspectives on ICT growth}

In order to gain a more thorough understanding of the factors affecting the provisioning of ICT infrastructure for video streaming in the UK, we decided to elicit the views of relevant industry stakeholders involved in that process. We augmented their views with those gathered from a short workshop and interviews with representatives from policy making groups. We used the results to identify what they consider to be the key drivers of the UK's ICT footprint growth. Video streaming was used as a specific case study but other energy demanding applications were also considered such as AI, blockchain and Internet of Things (IoT). We were also able to identify gaps and limitations within reporting and methods for studying the UK ICT footprint.

\subsection{Video streaming infrastructure and growth}

\subsubsection{Infrastructure and service location}

From the interviews, we found that capacity to enable video streaming was related to the need to position servers and content close to where videos are being delivered to the home of end users to reduce latency (be responsive) and alleviate burden on the network core. The physical and network location of data is important to manage load on the network, increase performance (reduce latency and jitter, provide sufficient bandwidth) and meet demand reliably (resilience). This includes decisions about the physical and network locations of data centres and content delivery networks (CDNs), such as the concept of edge computing (discussed by P2 and P13). This was described as a process of decentralisation (P11) where instead of favouring large central data centres, organisations were preferring to spread capacity to smaller facilities based on location. We were unable to find direct quantitative evidence to indicate the ICT footprint of transitioning from large scale data centres to edge computing beyond speculation by our sources, and none specifically within a UK context.

Other decisions related to how the data is processed through rendering and encoding into compressed data formats for transmission, and where such compute would happen on servers or end-user devices (P3, P6). Computation is also needed to compress video streams to be rendered at different resolutions or compression formats. P2 identified ongoing research on how content could be delivered in more efficient ways (i.e., unicasting or multicasting of the network data (P3, P7, P10), a longstanding challenge in the large-scale Internet . There were also decisions to be made about to what extent consumer devices were provided to end users to receive the video stream, such as routers and TV edge boxes commonly found in homes.

The decision-making processes described often included considerations on cost, power consumption, and sometimes carbon emissions and environmental impacts. In terms of sustainability, a specific tension was noted by P4 between the power usage of devices provided to consumers and when to replace them to maximise their life and amortise embodied GHG costs. This meant finding the right balance between extending the life of such products on sustainability grounds and need to replace equipment due to perceived risk due to inability to update software or incompatibility with the latest technologies employed in video streaming.

\subsubsection{Business growth drivers}

Changing and updating technology was identified as an important factor in growing and extending businesses. New technologies enabled a transformation of television watching and thus, the invention and popularisation of video streaming (P11: “what we see is just a rapid or relatively steady increase in video streaming being the sort of the viewing method of choice”). A complete transition from broadcast television was repeatedly mentioned (P3, P6, P15), and this appeared to be a key consideration in the expansion and development of the supporting infrastructure (i.e., how to handle this transition, associated audience sizes and, therefore, viewing peaks, which now manifest as a demand for concurrent video streams and personalisation of content choice).

In their search to be prepared for future technology developments, R\&D departments such as in organisations 1 and 3, experimented and developed proof-of-concepts and prototypes to demonstrate possible future trajectories to partner and customer organisations. This led to new types of services and products and growth to businesses. These ‘visions of the future’ were used to understand possible impacts on networks and user behaviour and shape organisation’s understanding on how future strategies around the evolution of network, facilities, devices, and services should be shaped. Such understandings of the future were embedded into goals, objectives, and attendant research projects in these businesses. This is to say, businesses were actively working towards how to support such transitions, and the developments that would be needed in their wake. 

In addition to the future planning, organisations (P3, P6, P13) were driven and responsive to the need to ensure high quality of service and availability to customers and end users. As P13 explained “the better you make it, the more it will get used. So it will drive an increase in usage and it will drive an increase in the amount of requirements that all these companies have”. Although organisations were only responsible for maintaining and improving part of the chain of services or infrastructure of video provision, they worked with other organisations in their supply chain to avoid issues such as poor latency, buffering or loss of service provision. When those collaborations were not in place it created difficulties. P4 commented how not knowing what other organisations plans for the future were (i.e., increasing their service or not to 4K streaming) made it harder to make decisions around service provision and carbon reduction, highlighting a need for better integration or communication between actors in the supply chain. 

The overarching narrative of business drivers was the competition and collaboration of different organisations who are a part of shaping and transforming video streaming possibilities in the UK. Some organisations had to collaborate with others in order to provide an infrastructure capable of providing an ever-expanding ecosystem of video streaming services in the UK, others were in direct competition with one another. The concept of ‘the attention economy’, referring to the limited attention capacity of the individual and the growth in content that lobbies for that attention, has proliferated alongside the growth in technologies and services \cite{nelson2020attention}. Attention economies becomes relevant when thinking of how those organisations involved in the creation of video content (i.e., television programmes, online videos) are caught up in the need to explore future opportunities and expand the capabilities of their services. With the growth in non-traditional video content (e.g., social media, live streaming), more traditional content producers are having to expand their services further to increase quality of experience, and provide new opportunities in video streaming (e.g., increasing the number of available video angles in sport). Such developments only spurred further competition between ‘more-traditional’ content providers as they sought to attract a bigger share of available audiences.

Developments such as immersive video and augmented reality (AR)/virtual reality (VR) (e.g., the metaverse) or higher speed mobile networks (5G) were considered as a potential threat in terms of an increase in data being streamed and computing required to run the services (P2, P13). For example, participants in immersive video for sports described how video footage was now produced from multiple viewpoints around stadiums. The system needs to have sufficient capacity to respond very quickly to avoid latency and render the required personal viewing experience as consumers switch viewpoints to maintain a high quality of experience. In other words, the capacity creates new features and services, which further enables more video streaming, growing overall demand for infrastructure. P13 also mentioned a potential new trend in retail (i.e., Amazon Go) streaming video from their brick-and-mortar shop in order to enable it to be processed in real-time in the cloud for monitoring shopper’s visits.

\subsubsection{Emissions related growth drivers}

Though consumer demand was identified as a business growth driver, it tightly links with emission growth drivers both in direct demand for energy to power the processing and transmission of video, but also for increasing the capacity and reach of the network to offer coverage, peaks in demand, redundancy for maintaining service in the event of service congestion or network and service failures, and managing uncertainty. Stakeholders (P6, P13) suggested that infrastructural development, and thus, the growth in video related ICT emissions was following consumer demand. This is to say, that as larger portions of the overall audience switch their viewing from broadcast to video streaming, the associated infrastructure also had to grow to meet this demand. P4 declared that \textit{“(video streaming) it’s the biggest part of our network traffic by a very long way”}. As noted above, video network traffic in 2022 constituted more than 65\% of all Internet traffic \cite{sandvine2023},  Sandvine also noted that “to keep up with these unprecedented demands, telcos have expedited network upgrades and buildouts” \cite{sandvine2023}.

Literature differs in claims regarding main drivers of change in terms of ICT emissions and appears to study such drivers within ICT system operation. Andrae and Edler \cite{andrae2015global} suggest that data centres will be a more significant contributor to the ICT footprint at 22-33\% (213 MtC02e) and stated that as users continue to adopt laptops and small screen devices instead of older PCs, they will contribute relatively less to the overall footprint (186 MtC02e). Their later work (e.g., \cite{andrae2019prediction}) anticipates the rise of AI and machine learning, VR and AR, as well as IoT will lead to big rises in data and traffic. There is limited data on the overall growth of AI, but the International Energy Agency IEA note Facebook’s application of AI alone has grown +150\% (compute for AI model training); +106\% (model inference) at the same time as +40\% in data centre growth \cite{iea2022}. IEA also note the emissions associated with PCs will continue to fall while those associated with smartphones will continue to rise. They predicted that data centre electricity consumption will likely rise with increasing volumes of data but that the increasing amount of renewables in the energy mix might lead to a reduction in overall footprint.

Video streaming is growing and constantly breaking ‘peak records’ (e.g., P3). This was particularly an issue when major events such as sport (i.e., final of a football tournament, P4) coincided with a gaming release or a large system update which also placed a concurrent demand on networks and data centres.  The size of major game titles grow in size in relation to the capability of consoles, TV resolutions and frame rates, and may be linked as franchises to major sporting events or movies. Such occasions put strain in the network and risk interfering with the quality of experience of consumers. Participants discussed (P4, P6, P10, P15) the need to manage and plan capacity for those peaks and suggested the need for events to be coordinated and for capacity planning. Although this was complicated as some of these events were global and thus the Internet related demand relating to this was harder to predict partially due to limited prior information from global sources. TalkTalk identified Amazon Prime’s successful bid for a package of Premier League matches as being a significant factor in hitting a new peak in their network traffic in February 2023; reaching 10.25 Terabytes per second. A similar delivery shift is expected with the introduction of, for example, Sky Glass (an Internet integrated TV) – users switching from the satellite service will instantly increase video related network traffic.

Because it is not clear when ICT demand peaks were going to occur, telecom organisations ensure networks have extra capacity to cope with anticipated peaks (P10 explained networks were designed to never hit full capacity). However once capacity to manage peak loads is in place, it is not taken away after a peak period has passed.  This growth in available capacity of the infrastructure provides headroom for new services to exploit. P2 mentioned that no-one would want to over-provision, but that it was perceived as being necessary. The constraint for over capacity is cost as explained by P11 \textit{“So if you are wasting server space that costs you”}.

The degree of over-provisioning in which a stakeholder is comfortable varied based on the nature of the ICT component they are operating. For example, Cloud data centres encounter frequent issues pertaining to designing their compute infrastructure to handle peak demand, resulting in an average system resource utilization of 12-18\% (where even globally leading companies such as Google typically only achieving 40-60\%). Hence, data centre over-provisioning is a widely adopted technique due to users vastly overestimating their compute resource requirements which can be effectively masked by technologies such as virtualisation. Such issues pertaining to addressing peak demand and extra capacity within ICT infrastructure is widely reported within the literature \cite{armbrust2010view}. Redundant infrastructures were also reported as necessary to provide resilience in the case of failure (i.e., extra equipment in case there is failure in parts of the system). This is seen as both a solution to potential technological challenges, but also an explicit selling point in order to attract risk averse customers. 

\subsubsection{More efficient upgrade fallacy}

Although newer and more efficient technology is constantly being developed and adopted, participants (P2, P4, P6), all working with delivering content via broadcast and networks, mentioned the need to maintain older and less efficient ‘legacy services’ in order to support older applications that were still in use, or meet ‘universal service’ obligations. Newer technology was seen to exist \textit{in addition to} pre-existing systems, \textit{adding} to related energy demand and emissions, rather than replacing it.  This is a factor that may not be considered when making claims for benefits of efficiency gain through newer generation hardware.

An example of this is the introduction of 5G alongside earlier generations in wireless communications technology \cite{carnstone2023}. 5G clearly providing lower direct energy per bit of data transmitted.  As we note from Cisco’s useful VNI survey \cite{cisco2020}, and our interviews with providers, earlier generations of technologies decline over time but get progressively harder to replace as they are held in place by 1) dependencies on the roll out of new infrastructure and sufficient coverage, 2) legacy technologies that build in dependencies to then current generations of hardware and software and their protocols, and 3) the need in some cases to maintain high levels of national coverage and availability (often incumbent service providers). Ultimately, as our participants discussed, once technology is rolled out it can have a potentially significant lifetime.  Estimates of overall net gains due to efficiency need to consider their potential lifetime and the base vs. operational energy demands, which we found from our literature review cannot be easily constructed due to incomplete ICT footprint data at sufficient granularity. Williams et al. note that embodied energy use and indirect energy use effects of 5G have been largely overlooked in most assessments of this kind \cite{williams2022energy}. 

In some instances, uptake is seen as an issue. P13 commented that despite more efficient technology for streaming and data centres being available (i.e., liquid cooling, more efficient processors) there was still low uptake because of increased cost and the lack of awareness, knowledge and skills concerning the ability to evaluate these technologies’ benefits. Government intervention and regulations were seen as being important to help stopping unsustainable practices (P10), to enable a level playing field and to keep the standards already in place due to EU regulations (for example, around data centres, P14).

\section{Discussion}

The GHG emissions associated with the ICT sector are currently on a par with those associated with global aviation \cite{FREITAG2021100340}\cite{wid2020}. They are also difficult to measure or even estimate. In our interviews, barriers mentioned in generating carbon emission data for organisations included lack of standardised way of measuring carbon emissions, difficulty in retrieving data on carbon emissions from supply chains, and no clear definition on who was responsible for the emissions of the chain of activities involved in delivering ICT services. When data is available it might not relate to the granularity of the information needed and the need to remain globally competitive was a driver of more local decisions.

The relationship between emissions and greater ICT use is complex because there are a number of interacting trends. For example, it can be argued that where ICT services replace more carbon intense existing services (e.g., video calling reducing the need to travel or an email replacing a physical letter) the net effect is a reduction in GHG emissions \cite{GESI2018}. This may be a convincing argument in developed countries with established existing procedures which are disrupted by new ICT services. But as we have seen, access to the Internet is far from universal and correcting that is likely to be a source of significant growth in ICT usage for decades to come. Furthermore, the increasing efficiency of equipment used to deliver ICT services may actually fuel the increased use. This ``rebound effect" is poorly understood when making changes to service provision \cite{FREITAG2021100340}. The lack of an ICT landscape model that captures all, or at least a significant proportion, of the sector's components and the relationships between them makes it unlikely that this understanding will be improved in the near future.

We then have to consider the actions of the many stakeholder involved in providing ICT services. As a whole, they are not actively making business decision which take account of the resulting effect on GHG emissions and even if they wanted to, they would struggle to find the necessary data. Between these stakeholders, there are also tensions. Those providing services to consumers do so, and create new services, to retain and attract customers because they are in competition with other businesses. Their decision places new requirements on infrastructure providers to ensure that they are capable of supporting those services. This necessitates a degree of cooperation between stakeholders but none are able to make business decisions with a clear understanding of the impact of those decisions on GHG emissions.

\section{Conclusion}

For a sector such as ICT which will likely see continued growth globally, this inability to accurately estimate current, and predict future, emissions is unsatisfactory when all other sectors are expected to reduce their emissions in the coming decades. What can be done to improve this situation? We see the need for developments in two areas. The first is that it is necessary to establish a thorough and complete understanding of the ICT landscape as a whole, not just that relating to video streaming, in such a manner that dependencies and responsibilities can be understood by all stakeholders. This requires the harnessing of industry knowledge and expertise which, as we have seen, can highlight important details of the underlying processes. The second is that there is a need for significantly more transparency in the reporting of energy usage and carbon emissions and there should be an expectation that this is done in a standardised manner. Furthermore, such disclosures should be decoupled from statements of ``green credentials" that companies routinely make. The exact format of energy usage and carbon emissions should be determined in consultation with appropriate industry experts and policy makers but should, for example, be broken down by country because of the importance of the carbon intensity of the electrical grid. These developments together should allow ICT stakeholders to better estimate the GHG emissions associated with their businesses and to better understand the consequences of the decisions they make.

\section*{Acknowledgment}

The authors would like to thank the following colleagues from the School of Computing \& Communications at Lancaster University for their assistance: Dr Carolynne Lord, Dr Marcia Smith and Dr Oliver Bates. This work was supported by the Engineering and Physical Sciences Research Council  (Fellowship number EP/V007092/1).

\bibliographystyle{IEEEtran}
\bibliography{JHRefs}

\end{document}